\documentclass{XrU2005}

\usepackage{graphicx}
\usepackage{natbib}
\bibpunct{(}{)}{;}{a}{}{,}
\def\hd{HD~152}
\def\cpd{CPD$-$41\degr}
\def\ngc{NGC~6231}
\def\sco{Sco~OB1}
\def\lg{$\log - \log$}
\def\lxlb{$L_\mathrm{X} - L_\mathrm{bol}$}
\def\xmm{XMM-{\it Newton}}
\def\epicmos{{\sc EPIC~MOS}}

\def\mos{{\sc MOS}}
\def\pn{pn}
\def\epic{{\sc EPIC}}

\def\cnts{cnt\,s$^{-1}$}

\def\ergscm{erg\,cm$^{-2}$\,s$^{-1}$}

\def\lg{$\log - \log$}
\def\farcs{\hbox{$.\!\!^{\prime\prime}$}}
\def\fs{\hbox{$.\!\!^{\rm s}$}}

\title{An exceptional X-ray view of the young open cluster NGC 6231: what XMM-{\it Newton} has taught us}
\author{H. Sana\thanks{FNRS Research Fellow}}
\author{E. Gosset\thanks{FNRS Research Associate}}
\author{G. Rauw$^\dagger$}
\author{J.-M. Vreux}
\affil{Institut d'Astrophysique et de G\'eophysique, Li\`ege University, All\'ee du 6 Ao\^ut 17,
Bat. B5c, B-4000 Li\`ege, Belgium}

\begin{document}


\maketitle

\begin{abstract}
Considered as the core of the Sco OB1 association, the young open cluster NGC 6231 harbours a rich O-type star population. In 2001, the XMM-{\it Newton} satellite targeted the cluster for a nominal duration of about 180 ks. Thanks to the detector sensitivity, the \epic\ cameras provided an unprecedented X-ray view of NGC~6231, revealing about 600 point-like sources. In this contribution, we review the main results that have been obtained thanks to this unprecedented data set. Concerning the O-type stars, we present the latest developments related to the so-called {\it canonical} $L_\mathrm{X}-L_\mathrm{bol}$ relation. The dispersion around this relation might actually be much smaller than previously thought. In our data set, the sole mechanism that yields a significant deviation from this scheme is wind interaction. It is also the sole mechanism that induces a significant variation of the early-type star X-ray flux. In a second part of this contribution, we probe the properties of the optically faint X-ray sources. Most of them are believed to be low mass pre-main sequence stars. Their analysis provides direct insight into the star formation history of the cluster. 
\end{abstract}

\section{Introduction}
Hot stars are known to be strong but soft X-ray emitters since the launch of the {\sc einstein} satellite in December 1978 \citep{HBG79,SFG79}. Although historically several hypotheses have been put forward to explain the origin of this X-ray emission, it is now commonly accepted that the latter is produced by shocks occurring within the denser layers of the wind. These shocks, believed to result from the growth of instabilities of the line driving mechanism, heat the wind material to temperatures of the order of ten million Kelvin, thus producing a substantial amount of soft X-ray emission. From the observational point of view, it was soon realized \citep[e.g.][]{HBG79, LoW80, PGR81} that the X-ray luminosities of hot stars were related to their bolometric luminosities through the so-called {\it canonical} \lxlb\ relation. In its generic form, this relation can be written as follows:
\begin{equation}
L_\mathrm{X} \approx 10^{-7} L_\mathrm{bol}.
\end{equation}
The X-ray luminosities of O-type stars seem however to present a large scatter  around this relation \citep[e.g.][]{BSD97}, typically of about a factor 2 to 3.

Beside the {\it intrinsic} emission of single stars, massive binaries are often more X-ray luminous compared to equivalent single stars \citep{Pol87, ChG91}. This additional X-ray emission is usually interpreted as the signature of a wind interaction resulting either from the collision of the winds of the two stars of the system or, if the wind from one component is much stronger than the other, from the interaction of the overwhelming wind with the secondary photosphere. This X-ray emission might further undergo phase-locked modulations, resulting from the variation of the absorption along the line of sight. In eccentric binaries, such phase-locked modulations might also result from the change of the shock strength due to the varying distance between the two stars.

Considered as the core of the \sco\ association, the young open cluster \ngc\ hosts a rich early-type star population of different sub-spectral types and luminosity classes. Located at a distance of only about 1.6~kpc, it constitutes an ideal target to study the X-ray properties of an homogeneous (in terms of age, reddening, chemical composition, ...) sample of OB-type stars. \ngc\ was thus chosen as the target of a 180~ks monitoring campaign performed in the framework of the Li\`ege project for the Optical Monitor consortium guaranteed time. Carried out within a 5-day period in September 2001, the campaign actually consisted of six separate pointings of 30~ks each. Nonetheless it forms one of the deepest X-ray views ever acquired of a young open cluster, but the particular schedule of the campaign further allows to study the variability of the X-ray properties of the objects on different time scales. The analysis of this remarkable data set is presented in a series of papers \citep{SGR05, SRN06, SRS06} to which we refer for further details. 

Beside the \xmm\ observations, we have also monitored all the O-type stars and some of the brightest B-type stars in the \epic\ field of view (fov) by means of high resolution optical spectroscopy (mainly acquired using the ESO spectrograph FEROS). This data set has allowed us to provide further constraints on the physical properties of the early-type stars in \ngc\ \citep{PhD}. In particular, we revised their multiplicity and re\-derived their spectral classification. Thanks to the quality of the FEROS data, we have detected the secondary signature for eight out of the nine O-type binaries in the \epic\ fov. Together with the existing photometry of the cluster \citep[e.g.][]{SBL98}, the tight constraints on the properties of the early-type stars provide a firm basis for the X-ray analysis. The complementarity between the X-ray and optical data further constitutes one of the strengths of the present work.

   \begin{figure}
   \centering
   \includegraphics[width=0.9\linewidth]{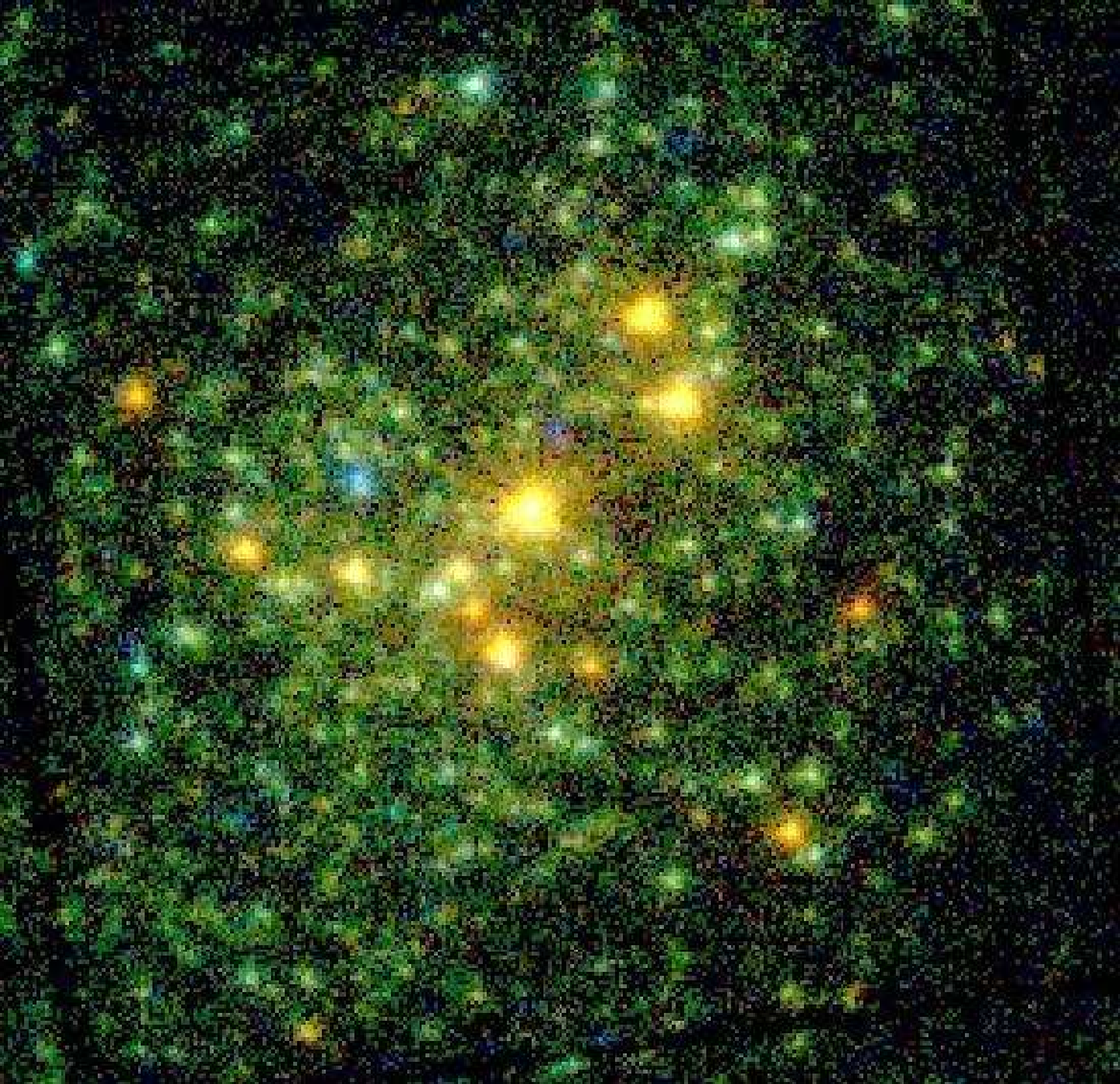}
   \caption{Combined EPIC-MOS three-color X-ray image  of the young open cluster \ngc, showing the central part of the fov. It approximately corresponds to MOS CCD\#1; the presented field is thus about $5\arcmin\times5\arcmin$. The different false colors correspond to different energy ranges: red: 0.5-1.0\,keV; green: 1.0-2.5\,keV; blue: 2.5-10.0\,keV. The color image is also available from the \xmm\ Image Gallery:  http://xmm.vilspa.esa.es/ external/xmm\_science/gallery/public/index.php}
   \label{fig: ngc6231}
   \end{figure}

\section{Observations and Data Reduction }\label{sect: obs}
\xmm\  performed the six successive exposures of our campaign during satellite revolutions 319 to 321. The fov was centered on the colliding wind binary \hd248 ($\alpha_{2000}=16^\mathrm{h} 54^\mathrm{m} 10$\fs06, $\delta_{2000}=-41$\degr49\arcmin 30 \farcs 1), in the core of the cluster. Position angles (PAs) were very similar through the six exposures. All three EPIC instruments were operated in the Full Frame mode together with the Thick Filter to reject UV/optical light. Due to the brightness of the cluster objects in the fov, the Optical Monitor was switched off throughout the campaign. 
	
The \epic\ Observation Data Files (ODFs) were processed using the XMM-Science Analysis System (SAS) v\,5.4.1 implemented on our computers in Li\`ege. We applied the {\it emproc} and {\it epproc} pipeline chains respectively to the \mos\ and \pn\ raw data to generate proper event list files. No indication of pile-up was found in the data. We then only considered  events with patterns 0-12 (resp. 0-4) for \mos\ (resp. \pn) instruments and we applied the filtering criterion XMMEA\_EM (resp. FLAG\,=\,0) as recommended by the Science Operation Centre (SOC) technical note XMM-PS-TN-43\,v3.0. For each pointing, we rejected periods affected by soft proton flares. For this purpose, we built light curves at energies above 10\,keV\footnote{Expressed in Pulse Invariant (PI) channel numbers and considering that 1 PI channel approximately corresponds to 1 eV, the adopted criterion is actually PI $>$ 10\,000.} and discarded high background observing periods on the basis of an empirically derived threshold (adopted as 0.2 and 1.0~\cnts\ for the \mos\ and \pn\ instruments respectively). The so-defined GTIs (Good Time Intervals) were used to produce  adequate X-ray event lists for each pointing from which we extracted images using x- and y-image bin sizes of 50 virtual pixels \footnote{Though the physical pixels of the \epicmos\ and \pn\ detectors have an extent on the sky of respectively 1\farcs1 and 4\farcs1, the virtual  pixels of the three instruments correspond to an extent 0\farcs05. The obtained images have thus a pixel size of 2\farcs5.}.

We finally combined the event lists obtained for all six pointings to improve the statistics of faint sources. For this purpose, we used the SAS task {\it merge}. For each \epic\ instrument, we included the event lists resulting from different pointings one by one. We also built merged event lists that combine the twelve \mos\  or the eighteen \epic\ event lists. The Attitude Files generated by the pipeline were merged using the same approach and we adopted, for handling the merged event lists, the Calibration Index File (CIF) and the ODF corresponding to the first pointing.

The total effective exposure times towards the cluster are, respectively for the \mos1, \mos2\ and \pn\ instruments, of 176.5, 175.0 and 147.5\,ks. Together with the high sensitivity of the \xmm\ observatory, the combination of the six pointings and of the three instruments provides one of the deepest X-ray views of a young open cluster. Fig.~\ref{fig: ngc6231} shows a three-colour image of \ngc\ and reveals a densely populated field with hundreds of point-like X-ray sources. We estimate our detection flux limit to lie between about $3\times10^{-15}$ and $1.5\times10^{-14}$\,\ergscm\ depending on the position on the detectors and on the source spectrum. 

For each source, we finally adopted a circular extraction region with a radius corresponding to half the distance to the nearest neighbouring X-ray source. Due to the crowded nature of the cluster core in the X-rays and to the limited spatial resolution of the \epic\ detectors, the background could not be evaluated in the immediate vicinity of the stars, but had to be taken from the very few source free regions. The details are given in \citet{SRN06}. We then used the appropriate redistribution matrix files ($rmf$) provided by the Science Operations Centre (SOC), according to the position of the considered source on the detectors. We built the corresponding ancillary response files ($arf$) using the {\sc arfgen} command of the SAS software. The spectra were finally binned to have at least 10 counts per bin. 

\begin{figure}
\centering
\includegraphics[width=0.9\linewidth]{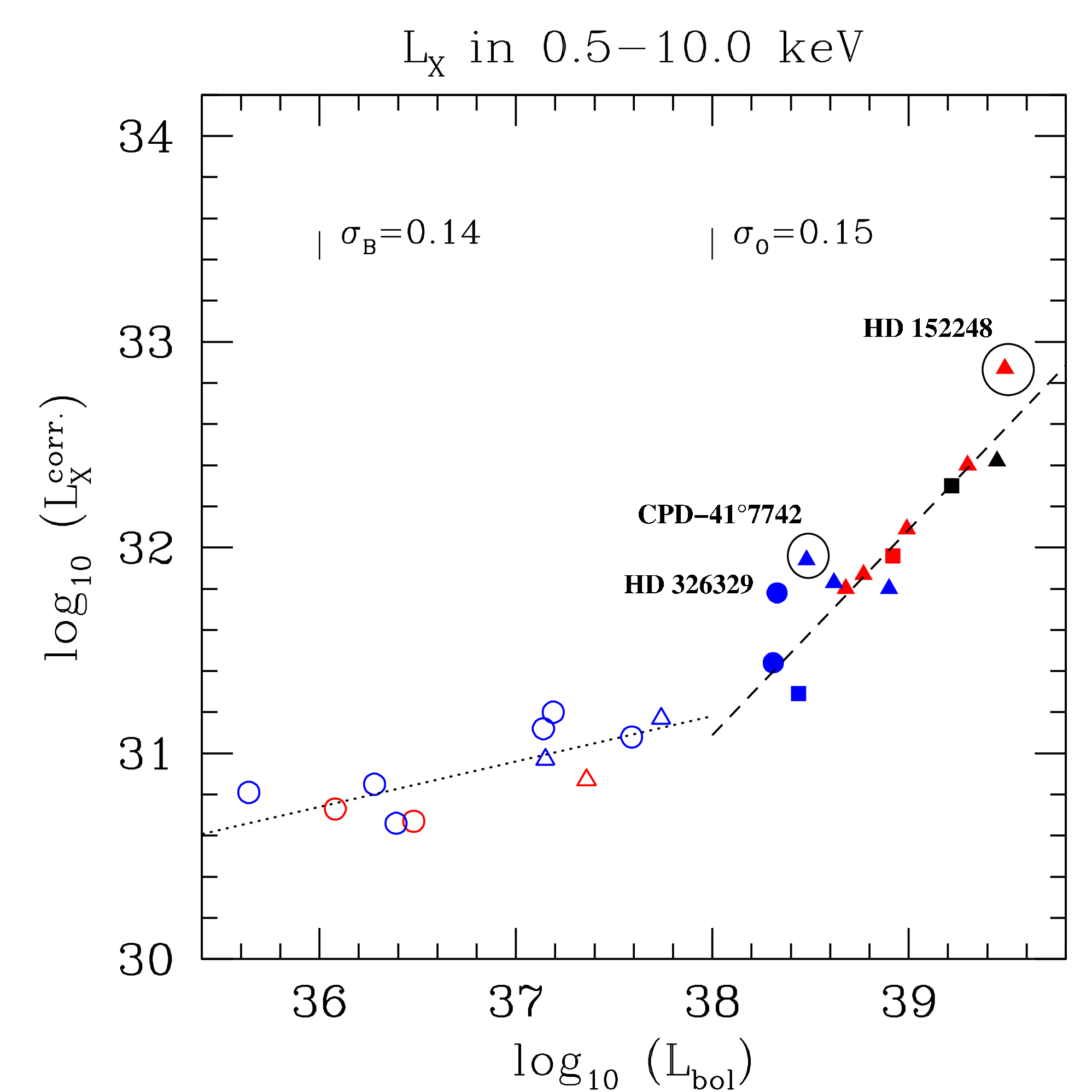}
\caption{ISM-absorption corrected X-ray luminosities in the 0.5-10.0~keV band plotted vs.\ bolometric luminosities. The different symbols indicate the different properties of the sources. Spectral type: O (filled symbols), B (open symbols). Luminosity class: supergiant (black), giant (red), main sequence (blue). Multiplicity: binary (triangles),  presumably single RV-variable star (squares), presumably single RV-constant star (circles). Best-fit linear relations in the \lg\ plane for O (Eq. \ref{eq: O}) and B (Eq. \ref{eq: B}) stars are indicated by the dashed and the dotted lines respectively. The two vertical bars in the upper part of the graph give the expected 1-$\sigma$ deviation for B ($\sigma_\mathrm{B}$) and O ($\sigma_\mathrm{O}$) type stars. \label{fig: lxlb}}
\end{figure}

\section{The OB star population}

Prior to our scientific analysis, we first went through a deep overview of the existing  literature and databases on \ngc\  and we made a census of the early-type star population in the observed fov. Our census mainly relies on selected published works \citep[see references in ][]{SGR05}, on the Catalog of Galactic OB Stars \citep{Reed03} and on the WEBDA\footnote{http://obswww.unige.ch/webda/} and SIMBAD\footnote{http://simbad.u-strasbg.fr/Simbad/} databases. This resulted in 108 objects among which 92 B-stars, 15 O-stars and one WR system (WR\,79). 

All the O-type stars were detected by \xmm\ as soft but strong sources, which allowed us to study the complete population rather than a subsample of it. On the other hand, only about 20\% of the B-type stars could be associated with an X-ray source. Using the {\sc xspec} software, we adjusted the X-ray spectra of the different sources associated with early-type objects. In doing so, we  used up to three-component thin thermal plasma models ({\tt mekal} models) allowing for a distinct local absorption column ({\tt wabs} model) for each component and accounting for additional absorption by the interstellar medium (ISM). The latter column was held fixed to a value computed from the reddening of the different objects, using the typical colours of \citet{SK82} and the gas to dust ratio of \citet{BSD78}. Using the best-fit models, we finally computed the X-ray luminosities, corrected for the ISM absorption column only. We also recomputed the bolometric luminosities, adopting our revised spectral classification and the bolometric correction scale of \citet{SK82}.
Fig.~\ref{fig: lxlb} presents the location of the different objects in the \lxlb\ diagram. Note that Fig.~\ref{fig: lxlb} is restricted to objects fitted with at least 2-temperature {\tt mekal} components. Reasons for this are given in \citet{SRN06}. 

\subsection{The O-type stars}

Focusing in a first step on the O-type stars, we note a clear linear relation in the \lg\ plane. However several points deviate significantly from this {\it canonical} relation. \hd248 \citep[O7.5III(f) + O7III(f),][]{SRG01} presents clear evidence for a wind-wind interaction that we traced both from the optical and X-ray domains. 2-D hydrodynamical simulations  of the collision further reasonably reproduce the observed modulations of the X-ray flux of the system, lending further support to this interpretation \citep{SSG04}. \cpd7742 \citep[O9V + B1.5V,][]{SHRG03} X-ray light curve also displays clear though apparently complex modulations of its X-ray flux, that we have however related to a wind interaction of a peculiar kind. In this O+B system, the dominant primary wind most probably crashes into the secondary surface. A simple phenomenological model of such a wind-photosphere interaction indeed reproduces pretty well the main features of the X-ray light curve \citep{SAR05}. It is thus clear that the two latter systems are not representative of the {\it intrinsic} X-ray emission of single O-type stars. Excluding these two points, a least-square linear fit yields:
 \begin{equation}
\log L_\mathrm{X} - \log L_\mathrm{bol} = -6.912 (\pm 0.153) \label{eq: O}
\end{equation} 
where the X-ray luminosity is given in the 0.5-10.0~keV band.
It is interesting to note that we also adjusted a power-law relation (thus in the form $\log L_\mathrm{X} = \Gamma \times \log L_\mathrm{bol} + K$). However,  as indicated by a $F_\chi$-test \citep[see e.g.][]{Bev69}, this additional parameter does not improve significantly the quality of the fit compared to the scaling law of Eq.~\ref{eq: O}. In the following, we thus decide to adopt the simplest form of the {\it canonical} relation, as quoted in Eq.~\ref{eq: O}. One of the most outstanding results lies however in the limited dispersion (of only about 40\%) of the X-ray luminosities around the best fit relation. Compared to the scatter (a factor of 2 to 3) observed by \citet{BSD97}, this suggests that the {\it intrinsic} X-ray emission of O-type stars might be much more constrained by the physical properties of the star than previously assumed. It is also remarkable that, within our sample, only the binaries present a significant modulation of their X-ray flux, suggesting thus a wind interaction origin. Excluding the case of HD~326329, probably contaminated by a neighbouring flaring source, extra emission produced in a wind interaction region is also the only mechanism that yields a significant deviation from the {\it canonical} \lxlb\ relation. All in all, our analysis suggests thus that the {\it intrinsic} X-ray emission from single O-type stars might be rather stable, both in terms of time-variability and of scattering compared to the mean \lxlb\ relation.

\begin{figure}
\centering
\includegraphics[width=0.9\linewidth]{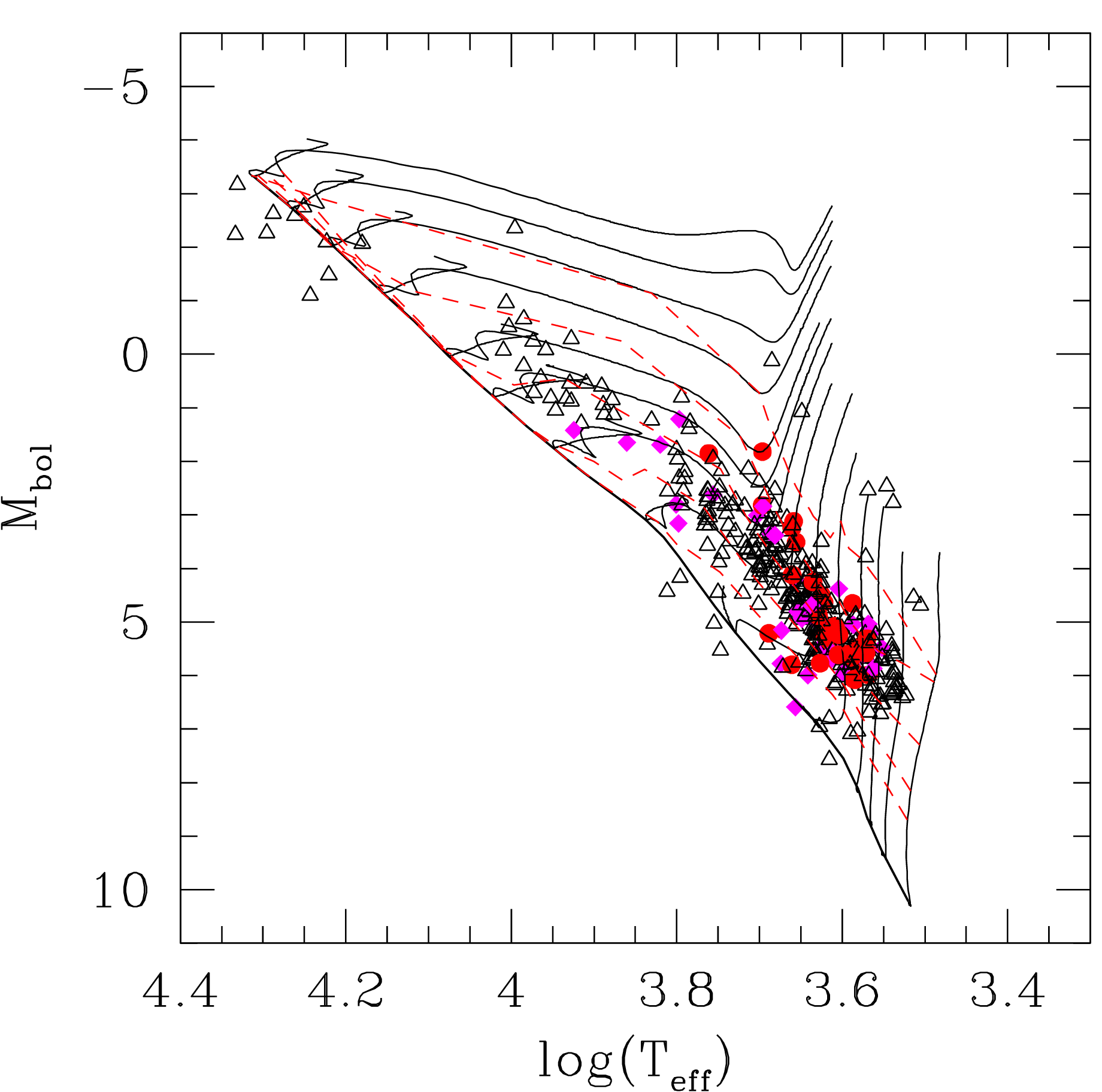}
\caption{Hertzsprung-Russell diagram of the EPIC sources with optical counterparts in the \citet{SBL98} catalogue. Evolutionary tracks from \citet{SDF00} for masses of 0.2, 0.3, 0.4, 0.5, 0.7, 1.0, 1.5, 2.0, 2.5, 3.0, 4.0, 5.0, 6.0 and 7.0\,M$_{\odot}$ are overplotted. Filled dots, filled diamonds and open triangles indicate respectively H$\alpha$ emitting stars, H$\alpha$ candidates and stars with no evidence for emission. The thick solid line shows the ZAMS, while the dashed lines correspond to isochrones for ages of 0.5, 1.5, 4.0, 10.0 and 20.0\,Myr.}
\label{fig: hrd}
\end{figure}
\subsection{The B-type stars}
Turning to B-type stars, their distribution in the $\log L_\mathrm{X} - \log L_\mathrm{bol}$ plane also suggests the presence of a linear relation, although as quoted above, only about 20\% of these objects are seen in the X-rays. Indeed the linear correlation coefficient is $r\sim0.75$, corresponding to a confidence level of 0.99 in favour of the presence of a correlation between $\log L_\mathrm{X}$ and $\log L_\mathrm{bol}$. A linear least-square fit yields: 
\begin{equation}
\log L_\mathrm{X} = 0.22 (\pm 0.06)  \log L_\mathrm{bol} + 22.8 (\pm 2.4)\label{eq: B}
\end{equation} 
However, we emphasize that the undetected B-type stars were not taken into account in the derivation of the present relation. The fact that about 80\% of them lie below our detection threshold suggests that either X-ray emission from B-type stars is not an {\it intrinsic} property of such stars or, at least, that it is not fully governed by their bolometric luminosities. In the latter case, we could only be detecting the upper envelope of a largely scattered distribution in such a way that, by coincidence, the detected sources suggest the presence of a linear relation. Beside these considerations, one may further note that about one third of the detected B sources present flaring-like activities during the time-span of our \xmm\ campaign. Their X-ray spectral properties are further reminiscent of those of PMS stars. It is therefore difficult to definitely conclude whether the detected X-ray emission is directly associated with the B-type stars or if it is rather produced by PMS stars either as the secondary component in a binary system or being located by chance along the same line of sight.

\begin{figure}
\centering
\includegraphics[width=0.9\linewidth]{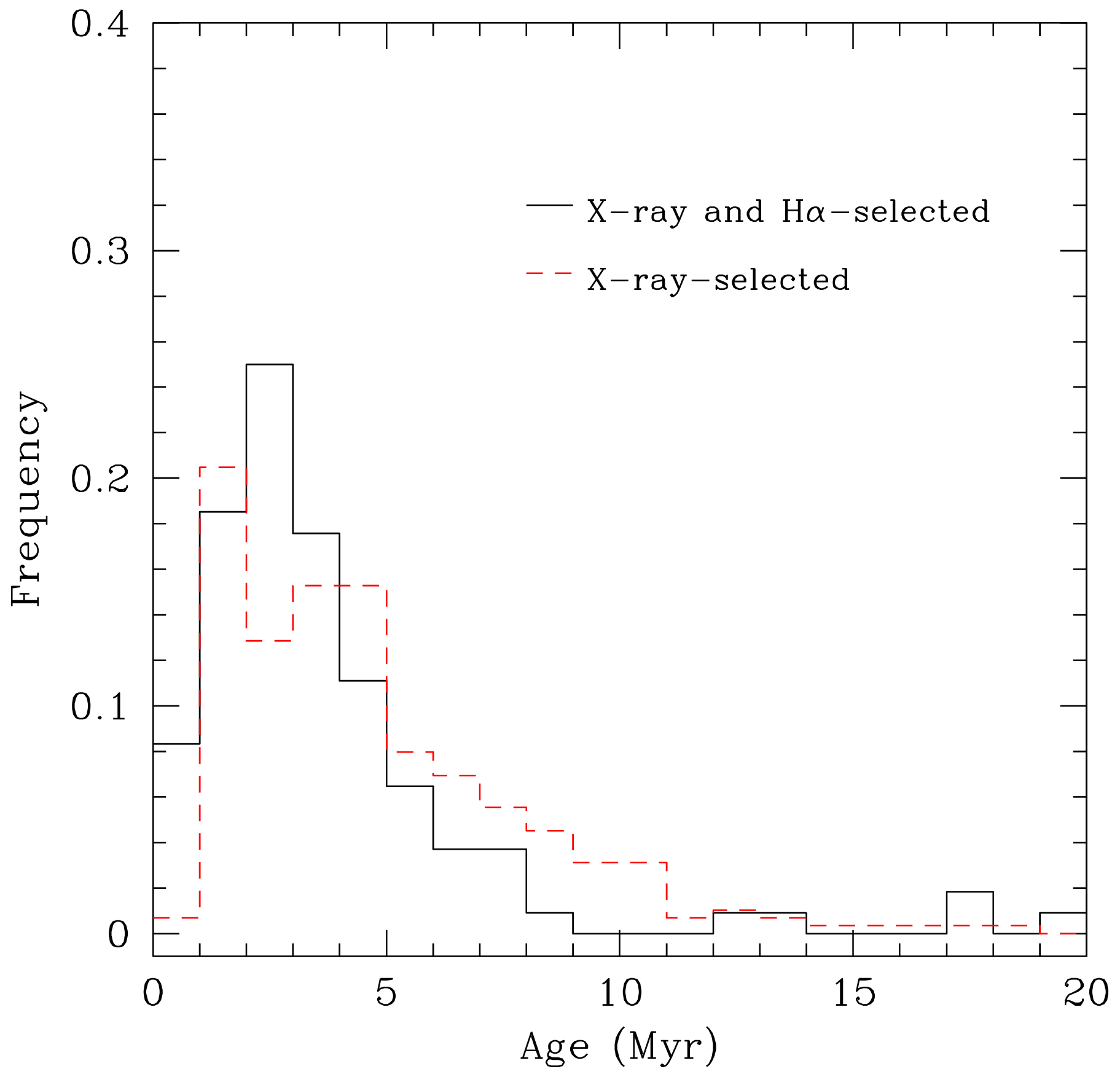}
\caption{Distribution of the ages of X-ray selected PMS objects as interpolated from the isochrones derived from the \citet{SDF00} evolutionary tracks. PMS candidates with $\Delta(R - H\alpha) \geq 0.12$ are indicated by the solid line, while those with $\Delta(R - H\alpha) < 0.12$ are indicated by the dashed line. The total numbers of objects with $\Delta(R - H\alpha) \geq 0.12$ and $\Delta(R - H\alpha) < 0.12$ are respectively 93 and 303.}
\label{fig: ages}
\end{figure}

\section{The optically faint X-ray sources}

Beside the early-type stars, the \epic\ images reveal several hundreds of additional point-like sources that cluster towards the core of \ngc, suggesting a physical link with the cluster. These are typically fainter but harder than the O-type stars \citep{SGR05}. As a first step, we compared the X-ray source list with different existing optical/IR catalogues, adopting a limited cross-correlation radius of 3\arcsec. As a result, almost 80\% of the X-ray sources in the \xmm\ fov could be associated with at least one optical/IR counterpart. In particular, we made use of an extended version of the UBV(RI)$_\mathrm{C}$~$H\alpha$ photometric catalogue of \citet{SBL98}. Roughly, the latter work covered a square field of 20\arcmin\ $\times$\ 20\arcmin\ that is inscribed within the 15\arcmin\ radius fov of the \epic\ cameras. \ngc\ being a quite young open cluster \citep[age$\sim$3-5~Myr,][]{BVF99}, we expect that the lower mass stars have not reached the ZAMS yet. Being probably still embedded in their natal clouds, they might suffer a heavy circumstellar absorption. X-ray emission is however one of the best selection criteria for these objects \citep{SB04} and the present \xmm\ image of \ngc\ might form one of the best censuses of the cluster PMS population so far. 
$H\alpha$ emission is a further well known property of the classical T Tauri stars. Among the 536 X-ray sources located within the field covered by the \citet{SBL98} catalogue, we indeed identified 93 candidates displaying a significant excess in the $R-H\alpha$ colour index \citep[adopted as $\Delta(R-H\alpha)\geq0.12$, see ][ for details]{SRS06}. 

\begin{figure}
\centering
\includegraphics[width=0.9\linewidth]{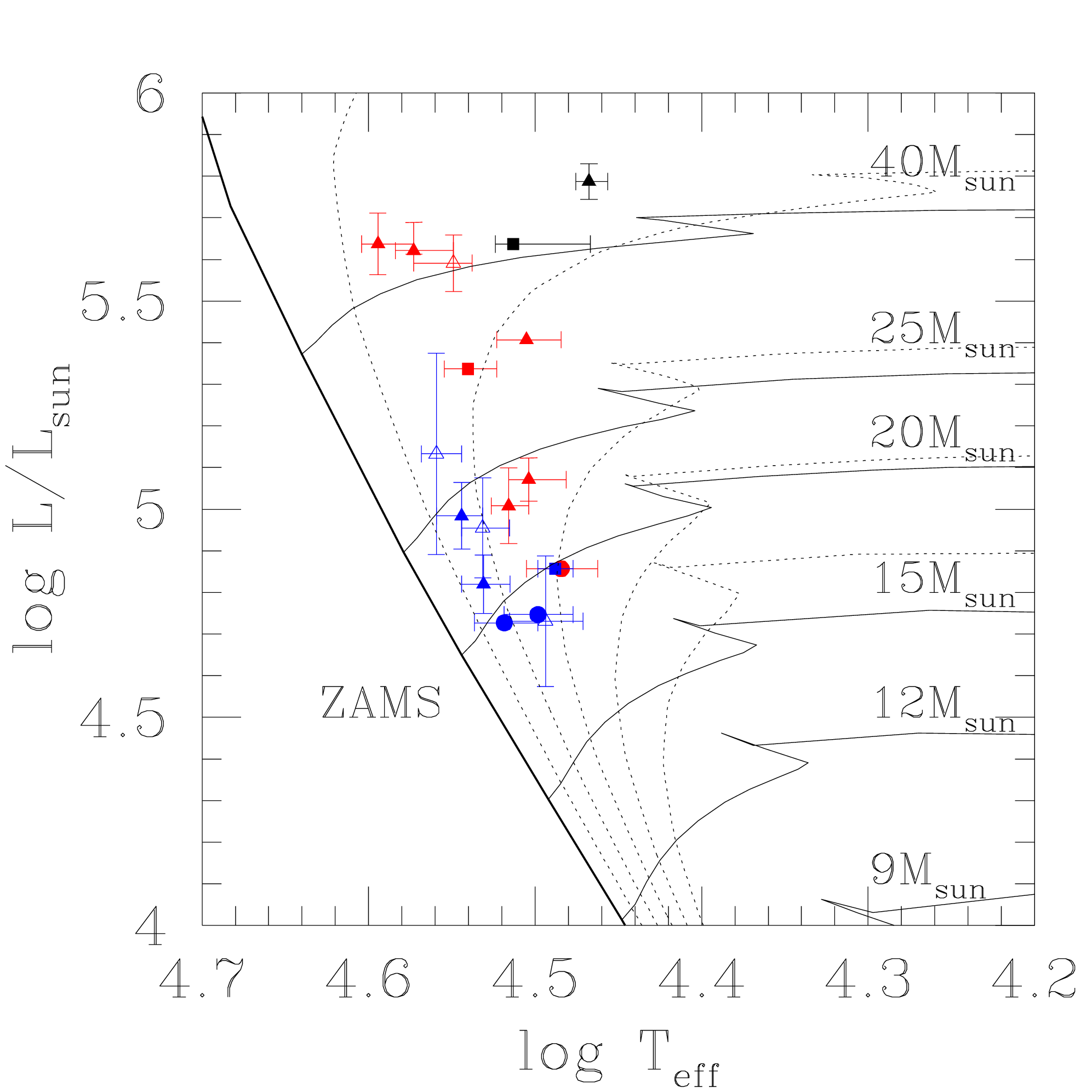}
\caption{Locations of the \ngc\ O-type stars in the H-R diagram. Symbol shapes and colours provide different indications on the object nature: black, supergiant; red, giant; blue, main sequence; triangle, belonging to a binary system (filled triangles indicate the heaviest star of the system while open triangles stand for the less massive companion); square, presumably single, RV-variable star; circle, presumably single, constant-RV star. The evolutionary tracks from \citet{SSM92} have been plotted (plain lines) together with isochrones (dotted lines) computed for ages ranging from 2 to 10 Myr with a step of 2 Myr. The figure appears in colour in the electronic version of this work.}
\label{fig: hr6231}
\end{figure}

Fig.~\ref{fig: hrd} presents the location of these optically faint X-ray sources in the Hertzsprung-Russell diagram. Most of them are indeed located above the ZAMS as expected for PMS objects. Comparing with PMS evolutionary tracks of \citet{SDF00}, it appears that most of them are low-mass stars ($M<2$~M$_\odot$) that started their formation about 2 to 4 Myr ago. The distribution of the ages of these X-ray selected PMS stars is presented in Fig.~\ref{fig: ages} and suggests that star formation in \ngc\ might have started at least 10~Myr ago at a relatively slow rate. The latter then slowly increased to culminate in a {\it starburst}-like event about 2 to 4 Myr ago, a period that also corresponds to the formation of the cluster massive stars (see Fig.~\ref{fig: hr6231}). 
It is further interesting to note that, neither in Figs. \ref{fig: hrd} nor \ref{fig: ages}, a clear difference appears between the H$\alpha$ emitting X-ray sources (probably classical T Tauri stars) and those that do not present evidence of such emission (probably weak-line T Tauri stars). No difference could also be found in their respective spatial distribution in the studied fov.

\section{Conclusions}
We briefly presented some of the main achievements of an \xmm\ monitoring campaign of the young open cluster \ngc\ in the \sco\ association. Of a nominal duration of 180~ks, it was actually split into 6 separate pointings spread over 5 days. Clearly it constitutes one of the deepest X-ray observations of a young open cluster and its particular scheduling allowed us to probe the variability of the detected sources on different
time scales. The \epic\ cameras revealed a crowded fov with more than 600 X-ray sources. The large majority of these could be identified with an optical/IR counterpart and their location in the H-R diagram suggests that most of them are actually low-mass ($M<2$~M$_\odot$) PMS stars. Their study seems to indicate that the star formation in \ngc\ was probably not an instantaneous event but might have started at least 10 Myr ago at a relatively slow rate. The bulk of the cluster stellar population has however started its formation during a starburst-like event about 2 to 4 Myr ago, an epoch during which the most massive stars of the cluster were also formed.

\xmm\ also detected the complete O-type star population in the \epic\ fov. Being strong but soft emitters, these objects clearly dominate the \epic\ images. Restraining our analysis to the 0.5-10.0~keV band, we derived a new value for the \lxlb\ {\it canonical} relation, expressed in the form of a scaling law. One of the most outstanding results is the limited dispersion of the O-type star X-ray luminosities around this new {\it canonical} relation compared to the one observed by previous studies. It is clear that the fact that we have been able, among others, to identify and reject the colliding wind binary systems from the fit has helped to reduce the scattering. The particular homogeneity of our sample, notably  in terms of age, metallicity and reddening, might have played a crucial role in this regard.

Finally, it is also worth to note that, within our sample, variability of the X-ray flux is only observed for probable colliding wind binaries. It is further the only mechanism that seems to produce strong deviations from the {\it canonical} relation. All in all, our analysis suggests thus that the {\it intrinsic} X-ray emission from single O-type stars might be much more stable than previously thought, both in terms of time-variability and of deviation from the mean \lxlb\ relation.

The comparison of the present results with others, derived in an homogeneous way from the study of other young open clusters, will therefore be of particular interest to probe the {\it universality} of the {\it canonical} relation. It could further help to investigate the influence of other important parameters, such as age and metallicity, which might crucially affect the {\it intrinsic} X-ray emission of O-type stars.

Finally we note that, compared to \xmm, future X-ray missions should definitely combine a high sensitivity with an increased spatial resolution. The latter is indeed crucial in very crowded fields such as the one studied in this contribution. We also note that, while dealing with such an extended data set, 12 months is a very short time. One can thus wonder whether the proprietary policy could not be extended to allow scientists to perform in-depth studies in good conditions.

\section*{Acknowledgments}

The present paper is based on data collected with {\it XMM-Newton}, an ESA Science Mission with instruments and contributions directly funded by ESA Member States and by the USA (NASA). The authors acknowledge support from the PRODEX XMM and Integral Projects, as well as contracts P4/05 and P5/36 `P\^ole d'Attraction Interuniversitaire' (Belgium). They are also greatly indebted towards the `Fonds National de la Recherche Scientifique' (Belgium) for multiple supports. HS is grateful to ESA, to the FNRS (Belgium) and to the `Patrimoine de l'Universit\'e de Li\`ege' for their contributions to the travel and accommodation expenses.

\bibliographystyle{XrU2005}
\bibliography{/datas6/XMM_CAT_PAPER/ngc6231_Xcat}

\end{document}